\documentclass[fleqn,usenatbib]{mnras}

\usepackage{newtxtext,newtxmath}
\usepackage[T1]{fontenc}
\usepackage{ae,aecompl}
\usepackage{etoolbox}
\makeatletter
\patchcmd\@combinedblfloats{\box\@outputbox}{\unvbox\@outputbox}{}{%
   \errmessage{\noexpand\@combinedblfloats could not be patched}%
}%
 \makeatother

\usepackage{graphicx}	
\usepackage{amsmath}	
\usepackage{amssymb}	
\usepackage{float}
\usepackage{gensymb}

\title[Study of phase differences between QPO harmonics]{A systematic study of the phase difference between QPO harmonics in black hole X-ray binaries}

\author[I. de Ruiter et al.]{
Iris de Ruiter,$^{1}$\thanks{E-mail: iris.deruiter@student.uva.nl}
Jakob van den Eijnden,$^{1}$
Adam Ingram$^{2,1}$
and Phil Uttley$^{1}$
\\
$^{1}$ Astronomical Institute, Anton Pannekoek, University of Amsterdam, Science Park 904, 1098 XH Amsterdam, The Netherlands\\
$^{2}$ Department of Physics, Astrophysics, University of Oxford, Denys Wilkinson Building, Keble Road, OX1 3RH Oxford, UK
}

\date{Accepted XXX. Received YYY; in original form ZZZ}

\pubyear{2019}

\begin{document}
\label{firstpage}
\pagerange{\pageref{firstpage}--\pageref{lastpage}}
\maketitle

\begin{abstract}
We perform a systematic study of the evolution of the waveform of black hole X-ray binary low-frequency QPOs, by measuring the phase difference between their fundamental and harmonic features. This phase difference has been studied previously for small number of QPO frequencies in individual sources. Here, we present a sample study spanning fourteen sources and a wide range of QPO frequencies. With an automated pipeline, we systematically fit power spectra and calculate phase differences from archival \textit{Rossi X-ray Timing Explorer (RXTE)} observations. We measure well-defined phase differences over a large range of QPO frequencies for most sources, demonstrating that a QPO for a given source and frequency has a persistent underlying waveform. This confirms the validity of recently developed spectral-timing methods performing phase resolved spectroscopy of the QPO. Furthermore, we evaluate the phase difference as a function of QPO frequency. For Type-B QPOs, we find that the phase difference stays constant with frequency for most sources. We propose a simple jet precession model to explain these constant Type-B QPO phase differences. The phase difference of the Type-C QPO is not constant but systematically evolves with QPO frequency, with the resulting relation being similar for a number of high inclination sources, but more variable for low-inclination sources. We discuss how the evolving phase difference can naturally arise in the framework of precession models for the Type-C QPO, by considering the contributions of a direct and reflected component to the QPO waveform.  
\end{abstract}

\begin{keywords}
X-rays: binaries
 -- black hole physics -- accretion, accretion discs
\end{keywords}



\section{Introduction}
Black hole X-ray binaries (BHXRBs) are stellar-mass black holes accreting from an orbiting stellar companion, that emit over most of the electromagnetic spectrum. The bulk of their luminosity is radiated in X-rays, emitted by the inner regions of the accretion flow, as most gravitational energy is liberated close to the accreting object \citep{first_disc_model}. X-ray observations of BHXRBs therefore provide a good mechanism for the study of the regions close to a black hole, which remain poorly understood even after decades of research.

An interesting feature that is observed in the X-ray light curves of accreting compact objects in certain spectral states are quasi-periodic oscillations (QPOs): regular but not completely periodic variations in the emission, appearing as a broadened peak in the X-ray variability power spectrum. A QPO in the light curve of a low-mass X-ray binary neutron star was first observed in 1985 by \citet{FirstQPO}. QPOs have subsequently been observed in many BHXRBs, especially after the launch of the X-ray timing observatory the \textit{Rossi X-ray Timing Explorer} ({\it RXTE}). In BHXRBs QPOs are observed at both high (50 to hundreds of Hz) and low (up to 10-20 Hz) frequencies.  Here we focus on low-frequency QPOs, where three different classes of low frequency QPOs have been identified, based on their frequency, Q-value, amplitude, noise and phase lag properties \citep{wijnands1999,remillard2002,ABC_table}. The Q-value is defined as the frequency of the QPO divided by its full-width half-maximum (FWHM), which is a measure for the width of the QPO peak in the power spectrum. In this paper, we will focus on the analysis of Type-B and Type-C QPOs. The main difference between these two types lies in the shape and amplitude of the broadband noise component in the power spectrum: this noise component has significantly lower amplitude ($\sim 1 \%$ rms vs $>10\%$ rms) when associated with Type-B QPOs than with Type-C QPOs. Most QPOs analyzed in this work are Type-C QPOs, since these are the most commonly observed, with a fundamental frequency ranging between 0.1 to 12 Hz, while the Type-B QPOs tend to cluster more between 4 and 6 Hz. We will ignore QPOs in the mHz and kHz range, which have also been observed \citep{highfreq, lowfreq2, highfreq2,lowfreq,lowfreq3}. QPO rms spectra indicate that the emitting region of Type-C QPOs appears to be the corona: an optically thin (optical depth $\tau \sim 1$) cloud near the black hole where cool disc photons are Compton up-scattered by energetic electrons \citep{qpo_from_cloud}. 

Different mechanisms have been suggested to explain the existence of QPOs. The fundamental distinction between these mechanisms lies in the origin of the observed variability, which can be divided into two categories: geometric or intrinsic. A geometric model assumes changes in (apparent) accretion geometry as the QPO origin, while the emission is intrinsically constant. An intrinsic model bases the QPO origin on modulations of intrinsic emission, e.g. driven by changes in accretion rate. As we will discuss below, recent evidence strongly favours geometric models, in particular the relativistic precession model \citep{rel_prec_model, rel_prec_model_2}. The key aspect of this model is the precession of a radially extended region of the hot inner flow of the accretion disk. If this inner flow -- which is hot and geometrically thick inside the inner radius of a truncated accretion disk -- has a spin that is misaligned with the black hole, it will start to precess due to (relativistic) Lense-Thirring torques. Changes in the observed solid angle of the flow and changes in Doppler boosting can then explain the quasi-periodic variations in flux. 

Recently, evidence for a general geometric QPO origin has been found in the inclination dependence of QPO amplitudes \citep{incl_dep_1, incl_dep_2} and the sign of lags between soft and hard photons \citep{incl_dep_3}. Recent research also shows that the reflection of photons from the inner flow of the disk depends on the phase of the QPO cycle \citep{Ingram, photon_refl_2, photon_refl_3}, which is a direct prediction of the relativistic precession model. While the relativistic precession model works well for Type-C QPOs, there are hints that Type-B QPOs also have a geometric origin \citep{incl_dep_1, geometric_typeb}. Finally, for the BHXRB GRS 1915$+$105, the energy-dependent behaviour of its low-frequency QPO can be explained by an extension of the solid-body precession model, where the inner flow precesses differentially \citep{differential_precession}.

For a better understanding of the origin of QPOs, spectral-timing methods provide a powerful approach. Such methods combine both the observed variability and spectral information to reach beyond the information available in each individually. For QPO studies, such methods naturally focus on studying spectral changes throughout the QPO oscillation cycle \citep{Ingram,geometric_typeb}. However, this approach fundamentally assumes that one can define a phase within the QPO cycle -- in other words, such methods assume the presence of a well-defined and stationary waveform throughout an observation. Interestingly, if such a waveform exists, it not only validates phase-resolved spectroscopy of the QPO but might also directly encode information about the origin of the QPO. 

Therefore, in this paper we set out to characterise QPO waveforms across a sample of BHXRBs over a wide range of their QPO frequencies. We use a statistical approach to determine whether a consistent phase relation between the QPO fundamental and harmonic exists and thereby (i) confirm whether a meaningful underlying waveform is present and (ii) search for signatures of the origin of the QPO harmonic. Fourteen different sources are analyzed, so the effects of inclination, QPO type and frequency on the waveform can be determined. To do so, the process of extracting the QPO properties from the original light curve is automated and the phase difference is calculated as a function of QPO frequency for a large sample of {\it RXTE} observations. 

\section{Methods}
\subsection{Rationale: a \textit{quasi}-periodic waveform?}
A periodic signal $x(t)$ consisting of two harmonics can, in the most general mathematical case, be written as:
\begin{equation}
	x(t) = A_f \cos(2\pi \nu_f t - \phi_f) + A_h \cos(2\pi\nu_h t - \phi_h)
   \label{eq:QPO_signal}
\end{equation}
where subscripts $f$ and $h$ correspond to fundamental and harmonic respectively, $A$ is the variability amplitude, $\nu$ the oscillation frequency ($\nu_h = 2\nu_f$), and $\phi$ is the phase offset. The absolute phase of the fundamental can be defined arbitrarily; in this paper, we set $\phi_f = 0$. If all amplitudes, frequencies, and phase-offsets are constant, this waveform will remain the same over time. 

\begin{figure}
\centering
	\includegraphics[width=\columnwidth]{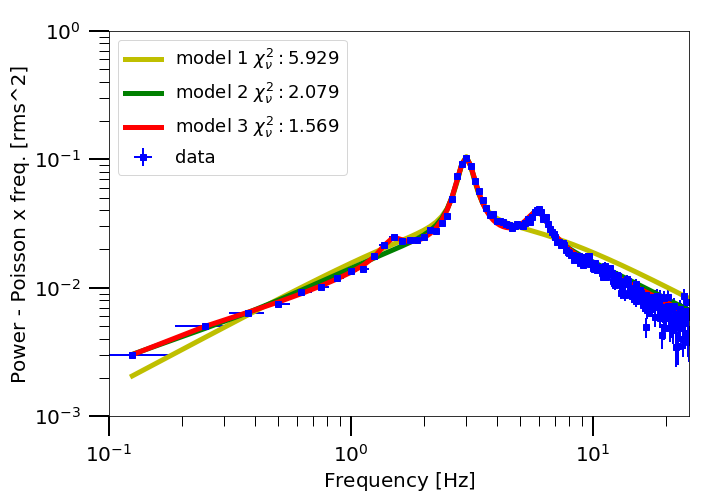}
    \caption{PSD for observation 20402-01-16-00 from GRS1915+105 in the 2-13 keV energy range. The Poisson noise, incorporated in all models, is subtracted from the power which is then multiplied by frequency by convention.  The reduced $\chi^2/\nu$-values for models one to three, which allow for increasingly more harmonic components, respectively are 5.929, 2.079 and 1.569}
    \label{fig:fit_example}
\end{figure}

The waveform of a QPO can be thought of in the same way as the deterministic waveform introduced above, consisting of multiple harmonic waves combining to a single waveform. The harmonic feature with the largest amplitude determines the main QPO frequency and is referred to as the \textit{fundamental} in this paper. Following common practice, the second harmonic, at twice the fundamental frequency, will simply be referred to as \textit{the harmonic} from here on. This harmonic can be much smaller in amplitude. An example of a QPO power spectrum in Fourier space is given in Fig. \ref{fig:fit_example}. From left to right, the spectrum shows three bumps: the sub-harmonic, fundamental and harmonic, respectively. By expressing the QPO waveform in terms of Equation \ref{eq:QPO_signal}, we will effectively ignore any subharmonic and higher order harmonic terms, approximating the QPO signal as just the sum of the fundamental and harmonic. While visible in Fig. \ref{fig:fit_example}, such features are rarely present in BHXRB observations. 

Considering a QPO following Equation \ref{eq:QPO_signal}, the quasi-periodic nature arises from the fact that the frequency/phase-offset and/or amplitude of each harmonic does not remain constant with time. However, for the waveform to be well-defined and stationary, each of these parameters should have a preferred value during the observation. The QPO frequency and amplitude have such a value, measurable from the power spectrum (see e.g. Fig. \ref{fig:fit_example}). The phase difference, $\Psi$, can not however be determined from the power spectrum. This phase difference is defined as:
\begin{equation}
   \Psi=[\phi_{\rm h}/2-\phi_{\rm f}] \mod \pi 
   \label{eq:phase_diff}
\end{equation}
This phase difference is the parameter of interest in this study, as we aim to test whether it has a preferred value in each observation; if so, together with the fundamental and harmonic frequency and amplitude, a well-defined waveform exists. If not, the waveform would change continuously and erratically during observations.

\begin{figure}
\centering
	\includegraphics[width=\columnwidth]{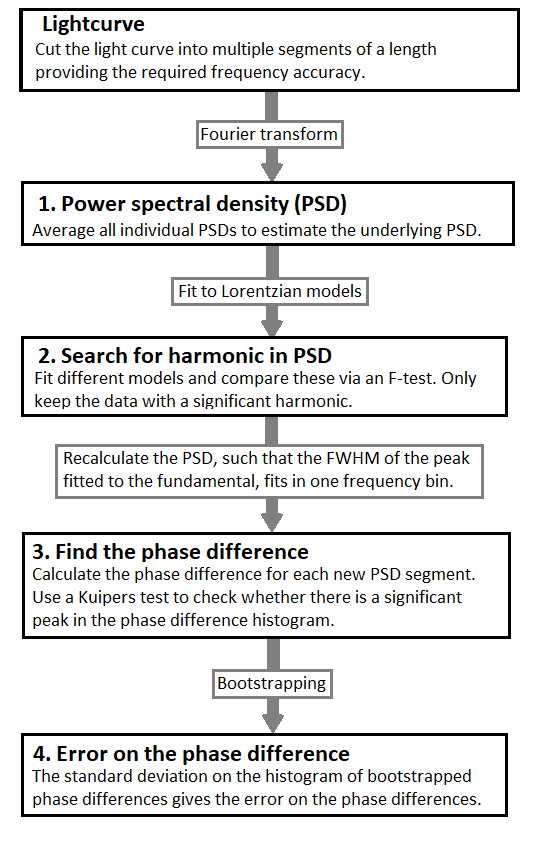}
    \caption{Schematic overview of the methods applied to measure the phase difference $\Psi$ between the QPO fundamental and harmonic. This method is based on \citet{Ingram}. The number of each step corresponds to the section explaining the step.}
    \label{fig:methods}
\end{figure}

In this section we will introduce each step of the method to calculate the phase difference in depth. Afterwards, Section 3 will describe the data we used to obtain our results and their automatic extraction from the {\it RXTE} data archive. A schematic summary of the pipeline, applied to each observation, can be found in Fig. \ref{fig:methods}, while each of the following subsections corresponds to one step in this figure. Schematically, the methodology is set-up in such a way that first, starting from individual light curves, the  power spectrum is created. This power spectrum is then fitted with different models to determine the QPO and harmonic parameters. Using these parameters, a Fourier analysis of individual small segments of the light curve is performed. This new power spectrum is finally used to find the phase difference, which is defined as $\Psi$ in Equation \ref{eq:phase_diff}. We stress that is not possible to calculate the phase difference directly from the cross spectrum; we refer the reader to the start of Section 3 in \citet{photon_refl_3} for a detailed explanation. 

The analysis method introduced here is based on the one developed by \cite{Ingram}, but generalized and automated in such a way that a single pipeline can be used for a large set of light curves. As stated, the next sections will provide a more detailed description of the steps of the analysis method.

\subsection{Creating the power spectrum}
\label{sec:creating_the_power_spectrum}
We calculate the power spectral density (PSD) in fractional rms normalization \citep{PSDnormal} for all individual light curves using a fast Fourier transform (FFT) algorithm. We average over $8$ s segments with a time step of $dt=1/128$ s, resulting in ensemble averaged power spectra in the $0.125$ to $64$ Hz frequency range.

\subsection{Fitting the power spectrum 
\label{sec:fitting_the_power_spectrum}}
We look for QPO fundamental and harmonic features in the PSDs by fitting with a multi-Lorentzian model \citep[following e.g.][]{PSDnormal}. We use two zero-centered Lorentzians for the broad band noise (BBN) \citep[following][]{differential_precession} and a narrow Lorentzian for each QPO harmonic. We consider a maximum of three harmonics (fundamental, first overtone and either second overtone or sub-harmonic) and use F-tests to determine how many harmonics to include in our best fitting model. We use a threshold of $p=0.01$ to determine whether the addition of a harmonic Lorentzian component is required by the data. Only observations with at least one significant overtone are useful for our analysis. An example of these three models fitted to a PSD is given in Fig. \ref{fig:fit_example} for {\it RXTE} ObsID 20402-01-16-00 (source GRS1915$+$105), which clearly shows the improvement of the fit as the model complexity increases. 

To be able to fit the power spectrum to the different models, an initial estimate for the QPO fundamental frequency is required. We obtain such an initial estimate by linearly rebinning the power spectrum by a factor three, averaging out statistical outliers. The frequency with the maximum jump in power (after subtracting the Poisson noise) in this rebinned power spectrum is our initial estimate for the QPO frequency. We use $P_{\rm noise}=2/\langle x \rangle$ where $\langle x \rangle$ is the mean count rate and the background is ignored;  e.g. \citet{1989fourier}; \citet{Uttley2014}). At this frequency, the power change is the most extreme and therefore represents an estimate of the QPO fundamental frequency at the resolution of the PSD. This method only yields the frequency of the sharpest peak and therefore isn't affected by other bumps in the PSD. Averaging over three data points ensures that there is a real peak building instead of one data point that is higher at random. A complete list of initial guesses and boundaries on fit parameters can be found in Appendix \ref{sec:app_parameter_guesses}.

\subsection{Phase difference}
\label{sec:phase_difference}
In order to measure the phase of the QPO fundamental and harmonic components, we first take the Fourier transform\footnote{Note that we use the same definition of the FFT as \citet{Ingram}, which is the conjugate of that used in the NumPy package in Python.} of many short segments of the light curve. Following \cite{Ingram}, we set the segment length for each observation to ensure that the QPO fundamental lies in a single frequency bin. The harmonic feature is therefore spread across two frequency bins. The phase of the fundamental for a single segment is calculated as the argument of the Fourier transform for that segment at the frequency bin containing the QPO fundamental. For the harmonic, we use the frequency bin at exactly twice the frequency of the bin used for the fundamental. We then calculate the phase difference using Equation \ref{eq:phase_diff}.  Note that throughout the remainder of this work we will consider the phase difference as a fraction of $\pi$ radians (which we denote $\psi/\pi$), to express it more clearly as a fraction of a cycle of possible values.

\begin{figure}
\centering
	\includegraphics[width=\columnwidth]{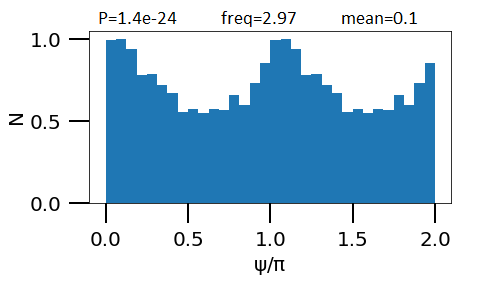}
    \caption{Phase difference histogram for observation 20402-01-16-00 from GRS~1915+105 in the 2-13 keV energy range. To show a clear peak we repeat the data from $\psi/\pi=0$ to $1$ to show two cycles. The probability that this distribution is randomly drawn from a uniform distribution is $1.4\cdot 10^{-24}$.}
    \label{fig:phase_diff}
\end{figure}

It is essential to calculate the phase difference for each segment individually, since the phase of the QPO fundamental and harmonic is random, and only their phase difference is constant. Averaging over the phases of the QPO fundamental and harmonic first and calculating the difference afterwards would result in a phase difference of zero (averaging over the difference of two random quantities). Since we calculate this phase difference for each segment, we can visualize the results in a histogram. An example of such a phase difference histogram is given in Fig. \ref{fig:phase_diff} for observation 20402-01-16-00 from GRS~1915+105 in the 2-13 keV energy range. We have repeated the $\psi/\pi = 0$ to $1$ interval twice in the figure, to show the peak clearly.  This histogram shows a clear peak around $\Psi/\pi=0.1$, which is repeated at $\Psi/\pi=1.1$. The histogram is normalized, such that the relative frequency of measurements at the largest peak is set to one. To check whether the histogram shows a significant peak, we use a Kuipers test \footnote{We use the Python implementation of the Kuipers test accessible via https://github.com/aarchiba/kuiper/blob/master/kuiper.py}. This Kuipers test is similar to a KS-test, but adapted for cyclic quantities, and returns the probability that the phase difference histogram is randomly drawn from a uniform distribution. Histograms are labeled significantly peaked using a significance level of $3\sigma$.

For phase difference histograms with a significant peak, the peak $\Psi_{\rm mean}$ is calculated by summing all distances for all $\Psi_m$ as follows:
\begin{equation}
    \begin{array}{l} \delta = \Psi_{\rm mean}-\Psi_m \hspace{1.85cm} \text{for} \hspace{0.3cm} \Psi_{\rm mean}-\Psi_m< \pi/2 \\ 
    \delta =\pi-(\Psi_{\rm mean}-\Psi_m) \hspace{1.1cm} \text{for} \hspace{0.3cm} \Psi_{\rm mean}-\Psi_m >\pi/2 \end{array}
    \label{eq:mean_distance}
\end{equation}
and minimizing the result with respect to $\Psi_{\rm mean}$. $\Psi_{\rm mean}$ has to be calculated in this way to deal with the cyclic nature of the phase difference. This can most easily be visualized by imagining a circle with circumference $2 \pi$ that describes all possible phases. Given two points on the circle, one for the fundamental and for the harmonic, we want to know the shortest distance between the two. The above calculation takes into account that this distance can be either clockwise or counterclockwise. This visualization of phases on a circle also explains why the phase differences range between 0 and $\pi$~rad instead of $2\pi$~rad. 

\subsection{Confidence interval on the phase difference}
\begin{figure}
\centering
	\includegraphics[width=\columnwidth]{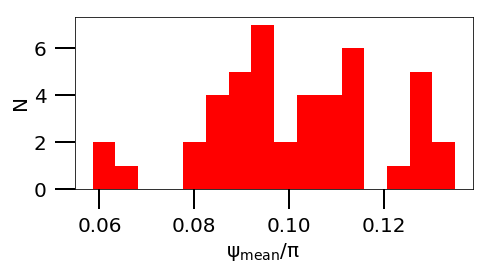}
    \caption{Bootstrapped phase difference histogram for observation 20402-01-16-00 from GRS1915+105 in the 2-13 keV energy range. From this distribution the standard deviation on the mean is calculated.}
    \label{fig:boot_hist}
\end{figure}

The final step in our analysis is to determine an error margin on the mean phase difference $\Psi_{\rm mean}$. Via bootstrapping, a statistical test that relies on repeated iterations of random sampling with replacement, we can assign an uncertainty to the phase difference. This is done by randomly drawing with replacement from the phase difference histogram. Using the mean of this new sequence and repeating this process multiple times, we build a new histogram that shows the bootstrapped mean phase differences. An example (using the same observation as shown in Figs. \ref{fig:fit_example} and \ref{fig:phase_diff}) is shown in Fig. \ref{fig:boot_hist}, which shows the distribution of inferred phase difference obtained from bootstrapping. The $1\sigma$ uncertainty of the phase difference is determined as the standard deviation of this bootstrapped distribution. For another example of how bootstrapping can be used to estimate parameter accuracy in spectral-timing studies, we refer the reader to  \cite{geometric_typeb}.

The number of bootstrap realisations of the data is determined using: $N_{\rm boot} = 50 - \rm int (\text{number of phase differences}/1000)$. This implies that the maximum number of bootstrapped mean phase differences used to determine the 1$\sigma$ uncertainty is 50 and will become smaller for long observations. The minimum number of bootstrap realisations is set to 15. This simple formula for the number of bootstrap realisations also optimizes the trade-off between a high number of bootstrap realisations and short computing time. To account for the cyclic property of the phase difference, the bootstrapped measurements $\Psi_{\rm mean, bootstrapped}/\pi \sim 0$ are shifted up by $1$ if  the actual value of $\Psi_{\rm mean}/\pi\sim 1$. This way $\Psi_{\rm mean, bootstrapped}/\pi=0.1$ is for example shifted to $1.1$ such that we obtain a distribution around $\Psi_{\rm mean}/\pi= 1$, without enormous gaps. The same principle applies to values of $\Psi_{\rm mean, bootstrapped}/\pi \sim 1$, which are shifted down by $1$ if the phase difference measured from the data itself $\sim 0$.

\section{{\it RXTE} observation sample}
For our analysis, we use the sample of {\it Rossi X-ray Timing Explorer} ({\it RXTE}) observations compiled by \citet{incl_dep_1} and further expanded and analyzed in \citet{incl_dep_3}. This sample contains 541 observations of 14 BHXRBs, of which 102 show Type-B and 439 show Type-C QPOs. We follow \citet{incl_dep_1} in their identification of QPO type, which is based on \citet{ABC_table}, and considers: the total RMS variability in the power spectrum ($\gtrsim 10\%$ for Type-C QPOs, $\sim 5-10\%$ for Type-B QPOs), the QPO frequency and its evolution during the outburst, the Q factor, and the shape of the noise component (flat-top-dominated for Type-C QPOs, red-noise for Type-B QPOs). The previous analyses of this data sample have shown that it is representative of QPO behaviour in BHXRBs and covers a range of source inclinations and possible QPO characteristics. A complete overview of the basic QPO and BBN characteristics of these observations, such as QPO frequencies, amplitudes and phase lags, can be found in the online materials of \citet{incl_dep_1} and \citet{incl_dep_3}. 
In Table \ref{table:sample}, we list the number of Type-C and Type-B observations in the initial sample per source. We reiterate that, since QPOs are not present in the soft state or in dim hard states, only a small fraction of RXTE observations of compact objects show statistically significant QPOs, depending on the observing strategy as well as the relatively rapid nature of the state transitions which show easily detectable QPOs. Therefore, the percentage of observations that contains QPOs differs per source and can range from less than one to $\sim 20$ per cent. Table \ref{table:sample} also shows the number of observations where we were able to successfully define a phase difference. The rest of the observations either lack a significant harmonic component or do not show a well-defined phase difference $\Psi$. The lack of significant harmonic could follow from short exposures or low fluxes, causing any harmonic features to become undetectable in the noise -- especially QPOs with fundamental frequency $\gtrsim 5$ Hz are prone to such effects, as their harmonics lie in lower signal-to-noise parts of the power spectrum. The lack of a measurable phase difference could also arise from short observations, resulting in few segments to calculate individual phase differences and therefore no significant peak in their distribution. Alternatively, a harmonic could simply be absent -- any model for its origin and properties should therefore be able to account for a weak power or even its absence. 
To extract light curves for every observation, we use the automated \textsc{chromos} pipeline\footnote{https://github.com/davidgardenier/chromos} \citep{gardenier18}. This pipeline automatically extracts light curves from raw \textit{RXTE} data in a specified energy range and with a pre-defined time resolution, independent of observation mode. We extract all light curves with a $1/128$ sec time resolution and a $2$--$13$ keV energy range. As the gain of \textit{RXTE}'s Proportional Counter Array (PCA) changes with time and the energy resolution depends on observing mode, \textsc{chromos} selects the absolute PCA channels most closely matching the energy range above for each individual observation. We also extract soft and hard light curves in the $2$--$7$ and $7$--$13$ keV ranges, matching those used in the QPO phase lag analysis of this sample by \citet{incl_dep_3}, to search for any energy dependence in the measured phase difference.

\begin{table}
\centering
\caption{The number of \textit{RXTE} sources per target in the sample, both \textit{before} and \textit{after} the analysis. The sample was originally constructed by \citet{incl_dep_1} and \citet{incl_dep_3}. Note that observations can be rejected in the analysis either if no harmonic is present or if no well-defined phase difference can be calculated.}
\label{table:sample}
\begin{tabular}{|l|c|c|c|c|c|}
\hline
\hline
Source   &  \multicolumn{2}{|c|}{Before} & \multicolumn{2}{|c|}{After} \\
 & C  & B    & C & B              \\ \hline
Swift J1753.5-0127 & 31 & -- & 21 & --           	\\ 
4U 1543-47  & 5 & 3 & 2& 2\\
XTE J1650-500  & 21 & 1 & 7 & 1        \\ 
GX 339-4    & 46 & 17 &15& 2\\ 
XTE J1752-223  & 3 & 1 &1 & 1\\
XTE J1817-330  & -- & 9 & -- & 5\\
XTE J1859+226  & 26 & 14 & 21& 7      	\\ 
XTE J1550-564  & 62 & 15 & 44& 8   \\ 
4U 1630-47  & 5 & -- & -- & -- \\
GRO J1655-40 & 27 & -- & 12 & --   	\\ 
H1743-322      & 100 & 36 &42 &14      	\\ 
GRS 1915+105   & 70 & -- & 68 & --   	\\ 
MAXI J1659-152 & 37 & 6 & 21 & --        	\\ 
XTE J1748-288  & 6 & -- & -- & -- \\ \hline
Total & 439 & 102 & 254 & 40\\ \hline \hline

\end{tabular}
\end{table}

\section{Results}
\begin{figure*}
	\includegraphics[width=\textwidth]{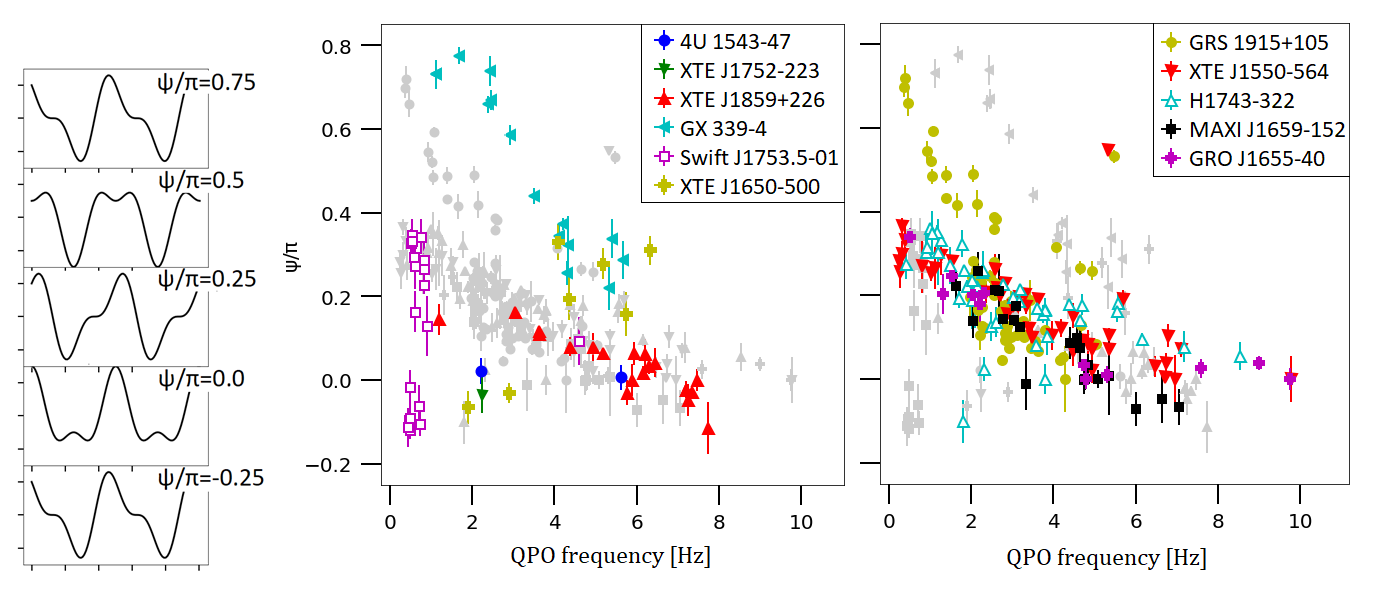}
    \caption{Phase difference between the fundamental and harmonic, as a function of the fundamental QPO frequency for Type-C QPOs. For datapoints with $\Psi/\pi>0.8$ we subtract $1.0$ such that these points are mapped to the $[-0.2 , 0]$ range. For visual clarity, only half of the sources are shown in color per panel. The other half of the sources are plotted in gray in the background, see the other panel. The left panel shows the low and undetermined inclination sources. The right panel shows the high inclination sources (see appendix \ref{sec:app_incl_dep}). The plots on the left side show the shape of the waveform corresponding to the indicated phase differences. Here $\nu_{\rm harm}= 2\,\nu_{\rm fund}$ $\rm Hz$ and \textbf{\textcolor{red}{$\rm A_{\rm harm}=0.5\rm A_{\rm fund}$.}} The plots have arbitrary time and intensity coordinates.}
    \label{fig:total_C_result}
\end{figure*}

\begin{figure}
	\includegraphics[width=\columnwidth]{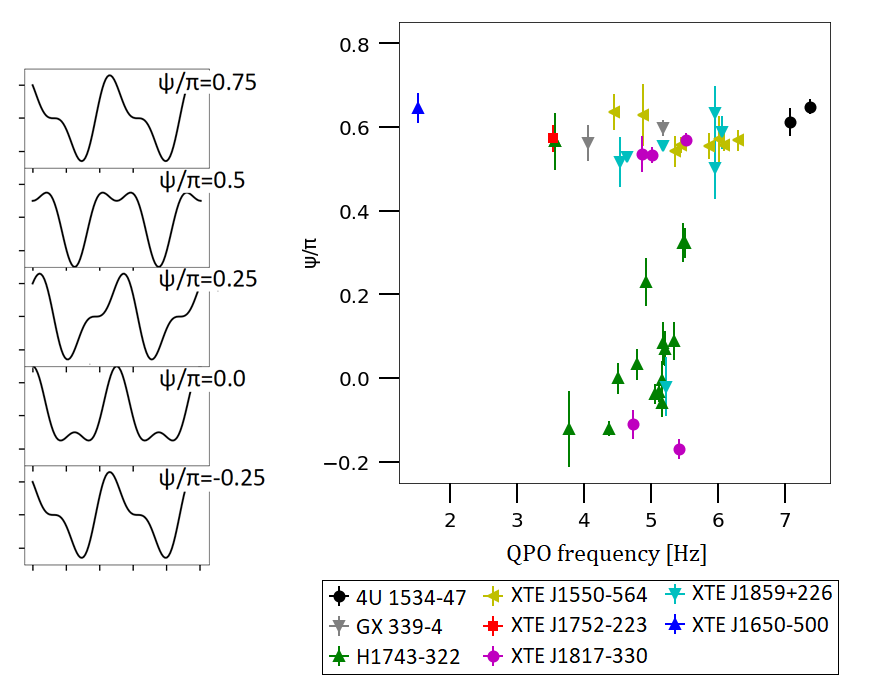}
    \caption{Phase difference between the fundamental and harmonic, as a function of the fundamental QPO frequency for Type-B QPOs. For datapoints with $\Psi/\pi>0.8$ we subtract $1.0$ such that these points are mapped to the $[-0.2 , 0]$ range. The plots on the left side show the shape of the waveform corresponding to the indicated phase differences, see caption Fig. \ref{fig:total_C_result} for details.}
    \label{fig:total_B_result}
\end{figure}

Figs. \ref{fig:total_C_result} and \ref{fig:total_B_result} show the phase difference between the fundamental and harmonic as a function of the fundamental QPO frequency for Type-C and Type-B QPOs respectively. To account for the cyclic property of the phase difference we subtract $1.0$ from data points with $\Psi/\pi>0.8$ such that these points are mapped to the $[-0.2 , 0]$ range. For clarity in Fig.~\ref{fig:total_C_result}, only half of the sources are shown in color per panel. The other half of the sources are plotted in gray in the background, and shown in color in the other panel.\\

From Fig. \ref{fig:total_C_result} it is clear that all sources have a well-defined phase difference ie. a well-defined QPO waveform, that evolves with QPO frequency. The phase difference decreases with frequency for all sources except for Swift~J1753.5-0127 (magenta open squares in the left panel) and XTE~J1650-500 (yellow crosses in the left panel). Furthermore most sources have a phase difference of $\Psi/\pi \sim 0.25-0.4$ at 0 Hz, and this phase difference decreases to around 0 at higher frequencies. Exceptions to this behavior are found in GX 339-4 (light blue leftwards pointing triangles in the left panel) and GRS 1915+105 (yellow dots in the right panel), which start at  $\Psi/\pi \sim 0.7-0.8$ at 0 Hz. Finally there are two data points that seem to be outliers at $\Psi/\pi \sim 0.55$ and $\sim 5.5$ Hz in the right panel (XTE J1550-564 and GRS 1915+105).\\

Type-B QPOs show a completely different behavior, as shown in Fig. \ref{fig:total_B_result}. In general the phase difference is constant with frequency at $\Psi/\pi \sim 0.55-0.6$. Obvious exceptions are found in H1743-322 (green upwards pointing triangles) and XTE J1817-330 (purple dots) that show a wide variety of phase differences.

The fact that sources show a systematic and well-defined phase difference for a given QPO frequency and QPO type, implies that for a given source's QPO type and frequency, there exists a simple underlying waveform of the QPO, that remains well-defined over each full observation. The presence of such an underlying waveform is a fundamental assumption of many recently developed QPO phase-resolving methods \citep{geometric_typeb,Ingram}: such methods aim to measure changes in the energy spectra as a function of phase within the QPO cycle, which evidently requires such a phase to exist throughout the analysed observation. 

For both the Type-B and Type-C QPOs, we made a preliminary analysis of the energy dependence of the phase difference by analysing light curves in the 2-7 and 7-13 keV bands. This analysis does not reveal any obvious dependence on energy. Therefore, we do not plot these extra energy bands in Figs. \ref{fig:total_C_result} and \ref{fig:total_B_result}. The lack of energy dependence might be explained by the broad nature of the two bands, averaging out any more subtle changes with energy. Alternatively, as discussed in Section \ref{sec:future}, it might be more fundamentally related to the origin of the QPO harmonic in relation to the energy range probed by {\it RXTE} PCA.  

\section{Discussion}

In this section, we will first discuss the search for any inclination dependence of our results. Afterwards, we introduce a new approach to identify QPO types based on the phase difference between the fundamental and harmonic. We then continue by discussing geometric models for the Type-B and Type-C QPO and whether these can explain the observed evolution of the QPO phase difference, and finish by recommending future steps for the observational and modeling side. 

\subsection{Waveform inclination dependence}
\label{sec:incdisc}

Different QPO properties -- such as the amplitude of Type-B and C QPOs \citep{incl_dep_1,incl_dep_2} and the Type-C fundamental energy-dependent phase lag \citep{incl_dep_3} -- have been shown to depend on the system binary orbit inclination, favoring a geometric QPO origin. The question therefore arises as to whether an inclination dependence can be found in the evolution of the QPO waveform.  Our sources show a variety of relations between the QPO frequency and the phase difference between harmonic and fundamental.  To show these differences more quantitatively, we fit a straight line to the data in Fig. \ref{fig:total_C_result} and \ref{fig:total_B_result} for each source individually. The fitted slopes and inclinations for each source are summarized in table \ref{table:line_fits_typeC} and \ref{table:line_fits_typeB} for Type-C and Type-B QPOs respectively in Appendix \ref{sec:app_incl_dep}. 

Although there is no single mapping of phase difference evolution to system inclination, it is worth noting two apparent patterns for the Type-C QPOs.  Firstly, there is a wide scatter in relations for low-inclination BHXRBs, with phase-difference either increasing or decreasing with frequency, or showing no obvious frequency-dependence. Meanwhile, the phase difference decreases with frequency for all high inclination BHXRBs, with data from four sources (i.e except GRS~1915+105) clustered around a similar relation.  The inclination of XTE J1859+226 is ill-defined (see appendix \ref{sec:app_incl_dep}) but its energy-dependent phase-lag evolution is consistent with a high inclination source \citep{incl_dep_3} and its phase difference evolution is also consistent with that of the majority of high-inclination sources.  Since the overall sample size is small and there is a wide range of patterns for the low-inclination sources, we cannot state that the differences are formally significant. However, they are certainly suggestive that there may be an inclination effect on the consistency of the phase difference evolution with frequency, with high inclination sources showing a more consistent pattern.

Although detailed modeling is beyond the scope of this paper, we can speculate how this behavior may be interpreted in the context of the precessing inner flow model \citep{rel_prec_model}. In this model, the precession period of the inner flow modulates the X-ray light curve via a number of mechanisms: variations during each precession cycle in the solid angle subtended by the inner flow, in Doppler boosting, and in the luminosity of seed photons from the disc intercepting the inner flow. The solid angle peaks at the point in the precession cycle when the flow is seen maximally face-on and Doppler boosting peaks when it is seen maximally edge-on \citep{rel_prec_model_2,ingram2015polarization}. The observed flux as a function of the angle between the flow rotation axis and the observer's line of sight is then a trade-off between these two effects. Crucially, if there are no more modulation effects, the observed flux is a monotonic function of this angle, such that the waveform we see from a given precessing inner flow system depends only on the angle between the black hole spin axis and the observer's line of sight.

This azimuthal symmetry is broken if the outer disc is misaligned with the black hole spin axis by some angle $\beta$ and the inner flow precesses around the spin axis maintaining a constant misalignment of $\beta$, as suggested in \citet{rel_prec_model}. The angle between the disc and flow rotation axes therefore varies over the precession cycle from a minimum of $0$ to a maximum of $2\beta$ (e.g. \citealt{rel_prec_model_2,ingram2015polarization}). As the angle between disc and flow changes, the number of seed photons from the disc intercepted by the flow changes, therefore varying the overall luminosity budget of the inner flow \citep{zycki2016energy, you2018x}. The QPO waveform now depends on the (polar) angle between the spin axis and the observer's line of sight \textit{and} on the azimuthal angle, defined for instance between the projections on the black hole equatorial plane of the observer's line of sight and the line of nodes between the black hole and disc rotation axes (e.g. see the schematic in \citealt{ingram2015polarization}). Perhaps for high inclinations, the solid angle and Doppler effects dominate over the variable seed photon effect. In this case, we would expect a roughly similar QPO waveform for a given geometry of the inner flow, with the waveform changing as the flow outer radius moves inwards. If the variable seed photon effect becomes more important for low inclination objects, then different low inclination sources would be expected to display different QPO waveforms, since we are presumably viewing them from a range of azimuthal angles.

For Type-B QPOs there is also no clear relation between inclination and phase difference. From Fig. \ref{fig:total_B_result} it seems that the phase difference is constant with frequency at around $\Psi/\pi \sim 0.6$ for 6 sources with Type-B QPOs, with the exceptions being H1743-322 and XTE J1817-330.

\subsection{A quantitative method to classify QPOs}
Examining the results of our analysis, shown in Figures \ref{fig:total_C_result} and \ref{fig:total_B_result}, two distinct types of QPO behaviour can be distinguished: on one hand, a large subset shows a decreasing phase difference with QPO frequency, as seen in Figure \ref{fig:total_C_result}. A smaller set of observations show QPOs in a small region of the QPO frequency-phase difference parameter space, where $\Psi/\pi > 0.5$ and $\nu>4$ Hz. This result represents a clear, qualitative method to classify QPOs into different types, based on only the QPO's properties and directly connected to its physical origin (which sets both the QPO frequency and phase difference).

An obvious question is then how this classification relates to the existing Type-B and Type-C classes of QPOs. The observations shown in Figures \ref{fig:total_C_result} and \ref{fig:total_B_result} are divided based on their prior separation into Type-C and Type-B QPOs, respectively. Therefore, our quantitative classification based on QPO properties alone almost perfectly overlaps with the existing classification, which relies on a more qualitative evaluation of a combination of QPO and BBN properties \citep[see e.g.][]{ABC_table,incl_dep_1}. 

\begin{figure}
\includegraphics[width=\columnwidth]{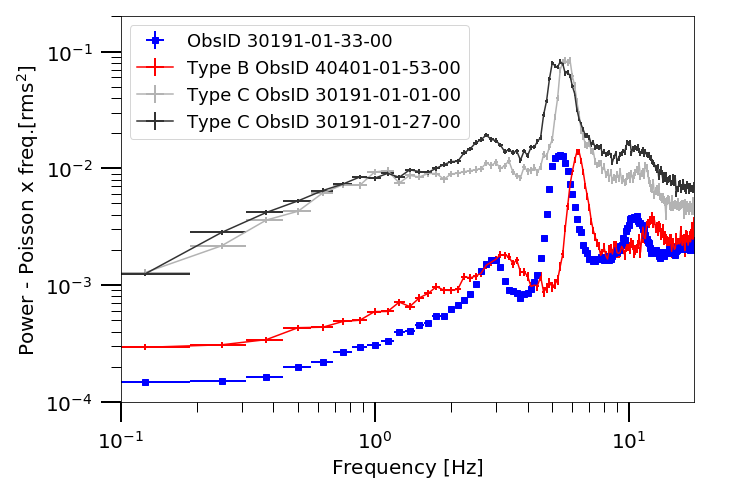}
\caption{PSD of the outlier XTE J1550-564 observation in the right panel of Fig. \ref{fig:total_C_result}. The outlier is ObsID 30191-01-33-00 from XTE J1550-564 which is shown in blue, a typical Type-B PSD with similar QPO frequency is shown in red, and two Type-C PSDs are shown in grey, all in the 2-13 keV energy range.}
\label{fig:outlier_spectrum_XTEJ1550}
\end{figure}

\begin{figure}
\includegraphics[width=\columnwidth]{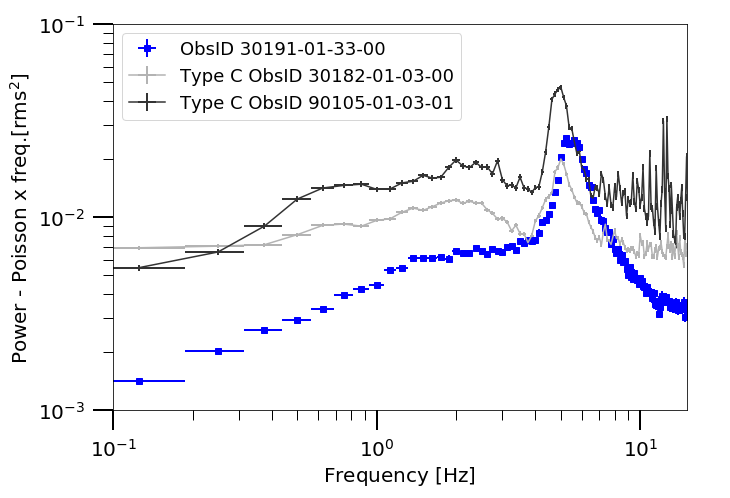}
\caption{PSD of the GRS~1915+105 outlier in the right panel of Fig. \ref{fig:total_C_result}. The PSD of ObsID 30402-01-11-00 from GRS~1915+105 is shown in blue and two typical Type-C PSDs of similar QPO frequency are shown in grey, all in the 2-13 keV energy range.}
\label{fig:outlier_spectrum_GRS1915}
\end{figure}

Considering Figure \ref{fig:total_C_result} in detail however, there are two observations identified as a Type-C QPO, which fit into our second category based on their phase difference (i.e. with $\Psi/\pi > 0.5$ and $\nu>4$ Hz). In Fig. \ref{fig:outlier_spectrum_XTEJ1550} and in Fig. \ref{fig:outlier_spectrum_GRS1915}, the power spectra of these two outliers are shown, together with power spectra of observations of the same source at similar QPO frequency. As shown in Fig. \ref{fig:outlier_spectrum_XTEJ1550}, the power spectrum of the outlier XTE J1550-564 observation is very similar to another Type-B QPO power spectrum (shown in red). It does not, however, look like the Type-C QPO power spectra shown as well in grey. Therefore, we conclude that this observation was most likely mis-classified as a Type-C QPO in \citet{incl_dep_1} and \citet{incl_dep_3}, showing the power of our more quantitative approach. Similarly, Fig. \ref{fig:outlier_spectrum_GRS1915} also appears different from the Type-C QPO power spectra shown: while the QPO itself has similar width and frequency, the noise level at lower and higher frequencies is significantly lower. While an unambiguous re-classification is difficult, it appears more similar to the Type-A QPO power spectra seen in for instance XTE J1550-564 \citep{wijnands1999} or MAXI J1535-571 \citep{stevens18}. 

\subsection{Interpretation of the Type-B phase difference}
\label{typeBmodel}
In this section we describe a simple geometric model that models the phase difference for Type-B QPOs. In our work we have not considered whether non-detections of a phase difference imply that in some cases there is no meaningful waveform (implicitly assuming that non-detections are due to signal to noise limitations alone), or the implications of the non-detections  of  harmonic  components  (which  would also require consideration of the upper limits on harmonic amplitudes). However if, in future studies, cases arise where there is clearly no meaningful waveform, or very weak harmonics, these would also need to be accounted for in any physical framework. The phase difference for the Type-B QPOs seems to be approximately constant at around $\psi/\pi\simeq 0.5$--$0.6$ for most sources. Recent studies suggest that Type-B QPOs could be related to a black hole with precessing jets \citep{geometric_typeb,typeBjet}. Therefore, we present a phenomenological model of a simple jet geometry in this section and analyze the resulting phase difference.\\

\begin{figure}
\includegraphics[width=\columnwidth]{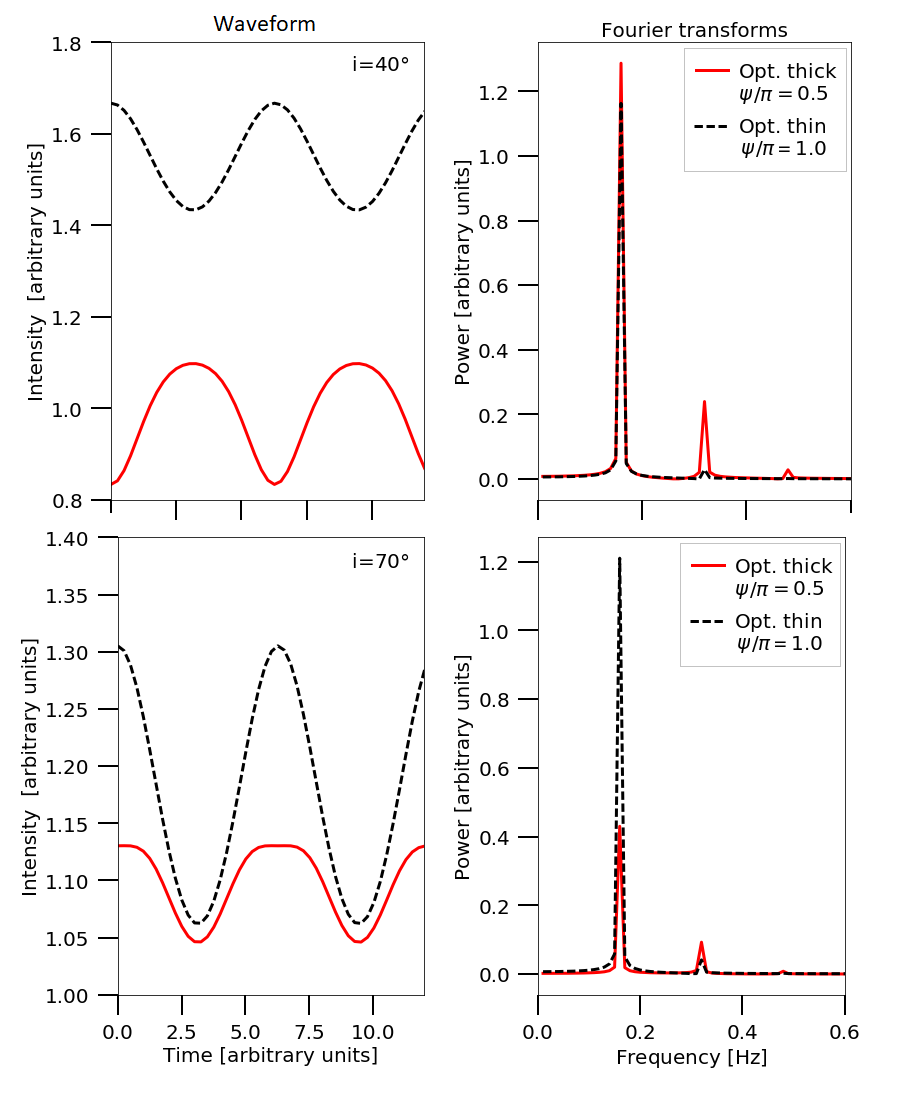}
\caption{Simulation of the waveform of a jet at an inclination of 40$\degree$ (top) and 70$\degree$ (bottom). The jet precesses with an angle of 10$\degree$ and has velocity $\beta=0.15$. The left plots show the actual observed waveform for an optically thick and thin jet. The right plots show the Fourier spectrum and the phase difference.}
\label{fig:jet_sim}
\end{figure}

In this model, we set up the jet simply as a narrow rotating cylinder (a `stick'), where the relativistic boosting and solid angle effects of an optically thick or thin emission region are taken into account (note that optically thin and thick refer to the X-ray properties and not the radio emission, and are therefore not related to the presence of either a steady compact jet or transient ejecta). This is a very simple representation of reality since the intrinsic angular dependence of emission from the jet may also play an important role. The observed luminosity for an optically thin jet can be approximated as \citep[e.g.][]{ellis1971}:
\begin{equation}
    S_{\rm observed}=S_{\rm emitted} \left( \frac{1}{\gamma[1-\beta \cos(\theta)]} \right)^4
\end{equation}
where $\beta$ is the jet speed in units of $c$ and $\gamma$ is the jet Lorentz factor. For an optically thick jet, the whole jet is not observed all the time, so the solid angle has to be taken into account:
\begin{equation}
    S_{\rm observed}=S_{\rm emitted} \left( \frac{1}{\gamma[1-\beta \cos(\theta)]} \right)^4 \sin(\theta)
\end{equation}
Here, $\theta$ is the angle between the line of sight of the observer and the jet. $\theta$ can simply be calculated by taking the dot product of the jet and observer vector

\begin{equation}
    \rm cos(\theta)= sin(\theta_{incl})\cdot sin(\theta_{jet})\cdot cos(\phi)+cos(\theta_{incl})\cdot cos(\theta_{jet}).
\end{equation}
Here, $\theta_{\rm jet}$ is the angle between the jet axis and the precession axis (i.e. the half opening angle of the precession cone), $\theta_{\rm incl}$ is the angle between the observer's line of sight and the precession axis and $\phi$ is the precession phase. In the following simulations, we set $\theta_{\rm jet}= 10\degree$, $\beta=0.15$ \citep{jet_params}, either $\theta_{\rm incl} = 40\degree$ or $\theta_{\rm incl} = 70\degree$, and $\phi$ varies from $0$ to $2 \pi$ during each precession cycle. We assume the intrinsically emitted luminosity is constant. Fig. \ref{fig:jet_sim} shows the result of this model: the left panels show the observed waveform for an optically thick (red line) and an optically thin (black dashed line) jet. The right panels shows the power spectrum of the waveform. For both inclinations the harmonic arises due to the harmonic content required to describe the non-sinusoidal, symmetric waveform, instead of a physical process occurring at twice the precession frequency. 

Furthermore, from this simulation it follows that independent of inclination the phase difference for an optically thick jet is $\Psi/\pi = 0.5$, close to the frequently-observed values. However, for the optically thin ($\rm i=40\degree$ and $\rm i=70\degree$) scenarios in Fig. \ref{fig:jet_sim} one might not find a significant harmonic component in an observation due to the low harmonic contribution.

In this model, only the simplest geometric effects of a `stick-like' optically-thick emitting jet are taken into account, so that a more realistic geometry (perhaps a composite of jet-like and disk-like), angle-dependent emissivity and general relativistic effects (likely significant in the part of the corona closer to the disk-plane) are not considered. Combining these effects with the simpler geometric effects described here could produce deviations from $\Psi/\pi = 0.5$ and perhaps explain the observed Type-B phase differences of $\Psi/\pi = 0.5-0.6$.

\subsection{Future steps and measurements}
\label{sec:future}

\subsubsection{Modeling}
The phase difference for the Type-C QPO systematically decreases with QPO frequency, for all except two sources. Such an evolving phase difference can be explained if the observed QPO consists of different contributions. For instance, in geometric precession models, the observed QPO waveform can be a combination of the direct emission from the precessing corona and reflected emission off the disk. The harmonic feature in the direct emission could be caused by general-relativistic effects distorting the waveform into a non-sinusoidal shape. The reflected component, on the other hand, could contain a feature at twice the QPO frequency because the precessing flow illuminates each disk azimuth twice per QPO cycle, with its back and front side, as implied by phase-resolved spectroscopy of H1732-322 \citep{photon_refl_3}. The two contributions to the net QPO waveform would likely not have the same phase difference. 

Mathematically, this idea corresponds to introducing two QPO waveforms, both with a different value of $\Psi$ that is constant as a function of QPO frequency. If the relative strength of these two waveforms varies with QPO frequency, the net phase difference of the observed waveform will evolve with QPO frequency as well. We can define a waveform $x_c(t)$ and $x_r(t)$ for the direct coronal and reflection QPO contributions, respectively, as:
\begin{equation}
	x_c(t) = A_{\rm f,c} \cos(2\pi \nu_{\rm f} t) + A_{\rm h,c} \cos(2\pi\nu_{\rm h} t - 2\Psi_{\rm c})
    \label{eq:corona}
\end{equation}
and 
\begin{equation}
	x_r(t) = A_{\rm f,r} \cos(2\pi \nu_{\rm f} t - \phi_{\rm r}) + A_{\rm h,r} \cos(2\pi\nu_{\rm h} t - 2\phi_{\rm r} - 2\Psi_{\rm r})
    \label{eq:ref}
\end{equation}
where the subscripts $\rm f$ and $\rm h$ refer to fundamental and harmonic, while the $\rm c$ and $\rm r$ correspond to coronal and reflected, respectively. Here, $\phi_{\rm r}$ is the phase difference between the fundamental components of the direct and reflected emission. The waveforms in Equations \ref{eq:corona} and \ref{eq:ref} are equivalent to Equation \ref{eq:QPO_signal}, using the definition of the phase difference $\Psi$ in Equation \ref{eq:phase_diff}. 

The superposition of the above two waveforms yields a full waveform with the well-defined phase difference $\Psi$, that depends on seven parameters of the original waveforms (namely all four amplitudes, the two phase differences, and the phase $\phi_{\rm r}$ in Equations \ref{eq:corona} and \ref{eq:ref}) in a non-trivial fashion. By fine-tuning any of these parameters (or more specifically, their dependence on QPO frequency), one can in principle reproduce any relation between $\Psi$ and QPO frequency. 

However, it is difficult to match physical scenarios for the QPO origin onto a single phase difference. For instance, both the phase difference between the fundamental and harmonic variation in the reflected flux, and the exact non-sinusoidal shape of the waveform of a precessing inner flow, are unknown and non-trivial to determine: they depend on for instance on the balance between modulation mechanisms discussed in section \ref{sec:incdisc}. Detailed examination of the different scenarios will require simulation of the relevant relativistic effects and radiative transfer, which will be an extensive effort and well beyond the scope of this work. In our work we have not considered whether non-detections of a phase difference imply that in some cases there is no meaningful waveform (implicitly assuming that non-detections are due to signal to noise limitations alone), or the implications of the non-detections  of  harmonic  components  (which  would also require consideration of the upper limits on harmonic amplitudes). However if, in future, cases arise where there is clearly no meaningful waveform, or very weak harmonics, these would also need to be accounted for in any physical framework.

\subsubsection{Observations}
We speculate that multiple mechanisms underlie the harmonic feature in the QPO, changing in relative strength as the QPO frequency and the outburst evolve. Both a reflection and a GR-based origin of the harmonic signal, discussed in the previous sections, could show a strong dependence on energy: reflection spectra are strongest below 2 keV, at the Fe K$\alpha$ line, and above ~15 keV (the Compton hump). Alternatively, the GR effects might dominate in the power law emission that originates from the innermost regions of the accretion flow. Therefore, if the harmonic contributions of one or both of these two possible mechanisms, this could be visible by considering the phase difference (evolution) in different energy ranges. 

In this work, we have considered the QPO waveform in a broad {\it RXTE} energy band between 2 and 13 keV. A preliminary test of any energy dependence, repeating the analysis for the 2-7 and 7-13 keV bands, did not reveal any clear effect of the choice of energy band. However, that is not particularly surprising considering the range of energies; given the {\it RXTE} response, both energy bands are dominated by the power law emission. In addition, the main feature of reflection, which might be related to the harmonic origin, in the {\it RXTE} band -- the iron line -- is located right at the division of the two bands. 

Instead of analysing the light curves in different energy bands, the phase difference as a function of energy can alternatively be calculated using simply the energy dependence of both the fundamental and harmonic phase lag (see the introduction to the method in \citealt{photon_refl_3}). Given RXTE's energy range and response, this approach is again not expected to show clear differences for {\it RXTE} observations. However, more novel observatories with different energy responses might reveal such an energy dependence; for instance, the recently employed {\it Neutron star Interior Composition ExploreR} ({\it NICER}) aboard the International Space Station is an X-ray spectral-timing instrument with a soft response (peaking below 2 keV) and excellent timing capabilities. Alternatively, the {\it Nuclear Spectroscopic Telescope ARray} ({\it NuSTAR}) can detect photons up to 79 keV and has been used previously to measure phase differences between the QPO fundamental and harmonic \citep{photon_refl_2}. Observing campaigns with these observatories tracking the QPO evolution and probing different parts of the reflection spectrum (beyond the reach of {\it RXTE} PCA), can further test whether multiple mechanisms underly the Type-C QPO harmonic. 

\section{Conclusions}
We have presented the first systematic analysis of the phase difference between the fundamental and harmonic of the QPOs from X-ray binaries. This phase difference gives important information about the physical processes that underlie the QPO emission mechanism, which is still unknown. We find that most of the studied X-ray binaries show a well-defined phase difference for a given QPO frequency and type, implying that they have well-defined waveforms. For Type-C QPOs we find that the phase difference decreases with QPO (fundamental) frequency except for Swift J1753.5-0127 and XTE J1650-500. For Type-B QPOs we find a completely different behavior with the phase difference in general being constant with frequency, at around $\Psi/\pi \sim 0.55-0.6$. Exceptions are found in H1743-322 and XTE J1817-330, that show a wide variety in phase differences.

A detailed analysis of the phase difference as a function of inclination does not yield a unique relation, although the data are suggestive that inclination plays a role in the consistency of the relation that is observed.  Type-C QPOs in high inclination sources show decreasing phase difference with frequency, with most sources clustered around a similar relation.  Low inclination sources show both increasing and decreasing phase difference vs. frequency relations.  For Type-B QPOs the behaviour is similar regardless of the inclination of the source. The different behaviors for Type-B and Type-C QPOs lead to a new, quantitative distinction between Type-B and C QPOs, improving on the more qualitative classification based on the overall power spectral shape and broadband noise strength. For the sources analyzed in this study we draw the conclusion that when $\Psi/\pi > 0.5$ rad and $\nu>4$ Hz, the QPO should not be classified as a Type-C. Using this new scheme, we have found two incorrectly classified QPOs in our sample, that were originally classified as Type-C QPOs by \citet{incl_dep_1} and \citet{incl_dep_3}.

To interpret the observed phase differences of the Type-B QPO, we propose a jet toy-model: the constant Type-B QPO phase difference observed in most sources can be explained by assuming that the QPO originates from the precession of an optically thick, X-ray emitting jet \citep{geometric_typeb,typeBjet}; the harmonic feature simply arises from the symmetric but non-sinusoidal nature of the resulting fundamental waveform. 

For the Type-C QPO phase difference, we observe a more scattered dependence on QPO frequency for low inclination sources. This larger scatter can be explained in the framework of geometric (Lense-Thirring) precession models. Such models, through the presence of both a direct and reflected component to the QPO waveform, might also be able to account for the evolution of the phase difference. However, detailed modeling is required to test our explanations of both the inclination-dependent scatter in, and the evolution of, the Type-C QPO phase difference. From the observational side, monitoring of the QPO with the X-ray observatories {\it NICER} and {\it NuSTAR} could further our understanding by searching for any energy dependence in the shape of the QPO waveform. 

\section*{Acknowledgements}
The authors thank the anonymous referee for constructive comments that improved the quality of this work, and are grateful to David Gardenier for sharing a copy of the \textsc{chromos} pipeline before public release. JvdE is supported by the Netherlands Organization for Scientific Research (NWO). AI is supported by the Royal Society. This research made use of Astropy, a community-developed core Python package for Astronomy \citep{astropy}. This research has made use of data and/or software provided by the High Energy Astrophysics Science Archive Research Center (HEASARC), which is a service of the Astrophysics Science Division at NASA/GSFC and the High Energy Astrophysics Division of the Smithsonian Astrophysical Observatory.




\appendix

\section{Data table \label{sec:app_data_table}}
The output produced by our pipeline, containing all data required to perform our analysis, is publicly available at the following url:  \href{https://github.com/jvandeneijnden/QPO-phase-differences}{https://github.com/jvandeneijnden/QPO-phase-differences}

\section{Lower and upper bounds and guess for fit parameters} \label{sec:app_parameter_guesses}
Initial guesses and lower and upper bounds for the fit parameters to models 1 to 3 as described in the methodology are shown in Table \ref{table:param}. Each model consists of a sum of Cauchy/Lorentz distributions, defined as:
\begin{equation}
P(\nu)=\frac{K\cdot (\sigma/2)}{(\nu-\nu_{\rm center})^2+(\sigma/2)^2}
\end{equation}
where K is the amplitude, $\sigma$ is the scale parameter that defines full width at half maximum and $\nu_{center}$ is the location parameter, specifying the location of the peak of the distribution. As stated before the two Lorentzians fitted to the BBN (described by $K_1$, $\sigma_1$, $K_2$ and $\sigma_2$) have a location parameter fixed at 0 Hz. $\nu_{\rm fund,0}$ is short for the initial QPO fundamental frequency guess. The C represents the constant power level that is added to the sum of Lorentzians. For model 1 only the BBN parameters ($K_1$, $\sigma_1$, $K_2$, $\sigma_2$) and the parameters for Lorentzian for the fundamental peak ($K_3$, $\sigma_3$, $\nu_{\rm center,3}$) are taken into account. For models 2 and 3 another Lorentzian is added each time, and the bounds on these Lorentzian parameters can be found in $K_4$, $\sigma_4$, $\nu_{\rm center,4}$ and $K_5$, $\sigma_5$, $\nu_{\rm center,5}$.

\begin{table}
\centering
\caption{Initial guesses and lower and upper bounds for the fit parameters to models 1, 2 and 3. $\nu_{\rm fund,0}$ is the initial QPO fundamental frequency guess.}
\label{table:param}
\begin{tabular}{l|l|l|l|}
                                     & Lower bound      & Guess            & Upper bound      \\ \hline
\multicolumn{1}{|l|}{$K_1$ [rms$^2/$Hz]}          & $10^{-4}$        & $10^{-3}$           & $\infty$         \\
\multicolumn{1}{|l|}{$\sigma_1$ [Hz]}     & 0                & $10^{2}$              & $\infty$         \\ \hline
\multicolumn{1}{|l|}{$K_2$ [rms$^2/$Hz]}          & $10^{-5}$        & $10^{-3}$            & $\infty$         \\ 
\multicolumn{1}{|l|}{$\sigma_2$ [Hz]}     & 0                & $10^{-1}$             & $\infty$         \\ \hline
\multicolumn{1}{|l|}{$K_3$ [rms$^2/$Hz]}          & 0                & $10^{-2}$            & $\infty$         \\ 
\multicolumn{1}{|l|}{$\sigma_3$ [Hz]}     & 0                & $1$                & $10$               \\ 
\multicolumn{1}{|l|}{$\nu_{\rm center,3}$ [Hz]} & $0.9\cdot \nu_{\rm fund,0}$  & $\nu_{\rm fund,0}$             & $1.1\cdot \nu_{\rm fund,0}$  \\ \hline 
\multicolumn{1}{|l|}{$K_4$ [rms$^2/$Hz]}          & 0                & $10^{-3}$            & $\infty$         \\ 
\multicolumn{1}{|l|}{$\sigma_4$ [Hz]}     & 0                & $1$                & $10$               \\ 
\multicolumn{1}{|l|}{$\nu_{\rm center,4}$ [Hz]} & $1.9 \cdot \nu_{\rm fund,0}$ & $2 \cdot \nu_{\rm fund,0}$   & $2.1 \cdot \nu_{\rm fund,0}$ \\ \hline
\multicolumn{1}{|l|}{$K_5$ [rms$^2/$Hz]}          & 0                & $10^{-2}$             & $\infty$         \\ 
\multicolumn{1}{|l|}{$\sigma_5$ [Hz]}     & 0                & $1$                & $10$               \\ 
\multicolumn{1}{|l|}{$\nu_{\rm center,5}$ [Hz]} & $0.2 \cdot \nu_{\rm fund,0}$ & $0.5 \cdot \nu_{\rm fund,0}$ & $0.8 \cdot \nu_{\rm fund,0}$ \\ \hline
\multicolumn{1}{|l|}{C [rms$^2/$Hz]}              & 0                & $10^{-3}$            & max power        \\ \hline
\end{tabular}
\end{table}

\section{Inclination dependence of the phase difference} \label{sec:app_incl_dep}
Tables \ref{table:line_fits_typeC} and \ref{table:line_fits_typeB} show the results of the straight line fits to Fig. \ref{fig:total_C_result} and Fig. \ref{fig:total_B_result} and inclination for the Type-B and C QPOs respectively. The table shows number of data points used for the fit (data points), the slope of the fit, the fit intercept with the y-axis, the inclination of source, the inclination sample the source belongs to and the reference for the inclination data. The errorbars on the slope and intercept represent a $1\sigma$ interval. For some sources only 2 data points were available, so the straight line fit has zero error. Sources with only one data point aren't included in this table.

\begin{table*}
\centering
\caption{Slope of straight line fitted to the phase difference vs frequency plots for Type-C QPOs. Two outliers for GRS~1915+105 and XTE~J1550-564 are disregarded. 4U~1543-47 and XTE~J1752-223 have only one or two data points such that fitting a straight line makes no sense. Inclination sample from \protect\cite{incl_dep_1}. References for the specific inclinations: [1]: \citet{incl_SWIFT_J1753}, 
[2]: \citet{incl_4U1543},
[3]: \citet{incl_XTEJ_1650}, [4]: \citet{incl_GX_339_1,incl_GX_339_2}, 
[5]: \citet{incl_XTEJ1752}, 
[6]: \citet{incl_XTEJ_1859}, [7]: \citet{incl_XTEJ_1550}, [8]: \citet{incl_GROJ_1655}, [9]: \citet{incl_H1743} and [10]: \citet{incl_GRS_1915}}
\label{table:line_fits_typeC}
\begin{tabular}{|l|l|l|l|l|l|l|}
\hline
Source        & Data points&  Fit slope  & Fit intercept  &Inclination                 & Sample & Ref.\\ \hline
Swift J1753.5-0127 &21  & $0.075\pm 0.092$ & $0.026 \pm 0.073$ & $\sim 40-55$                & Low	& [1]	\\ 
4U 1543-47       &2   & $-0.004 \pm 0.0$ & $0.033 \pm 0.0$  & $20.7 \pm 1.5 \degree$      & Low & [2]\\

XTE J1650-500    &7   & $0.093\pm 0.025$ & $-0.250 \pm 0.107$ & $> 47 \degree$              & Low	& [3]\\ 
GX 339-4         &15  & $-0.142\pm0.013$ & $0.975\pm 0.044$   & $40 \leq i \leq 60 \degree$ & Low	& [4]\\ 

XTE J1752-223    &1   & -                & -          & $\leq 49 \degree$           & Low & [5]\\ \hline

XTE J1859+226    &21  & $-0.036\pm 0.002$& $0.0244\pm 0.014$ & $\geq 60 \degree$           & - 		& [6]	\\ \hline
XTE J1550-564    &43  & $-0.043\pm 0.003$& $0.319\pm 0.012$ & $74.7 \pm 3.8 \degree$      & High 	& [7]\\ 
GRO J1655-40     &12  & $-0.033\pm 0.007$& $0.264\pm 0.033$ & $70.2 \pm 1 \degree$        & High	& [8]	\\ 
H1743-322        &42  & $-0.046\pm 0.005$& $0.348\pm 0.017$ & $75 \pm 3 \degree$          & High	& [9]	\\ 
GRS 1915+105     &67  & $-0.107\pm 0.013$& $0.497\pm 0.040$ & $70 \pm 2 \degree$          & High	& [10]	\\ 
MAXI J1659-152   &21  & $-0.064\pm 0.006$& $0.346\pm 0.024$ & -	                       & High   & -	\\ \hline
\end{tabular}
\end{table*}

\begin{table*}
\centering
\caption{Slope of straight line fitted to the phase difference vs frequency plots for Type-B QPOs. Inclination sample from \protect\cite{incl_dep_1}. References for the specific inclinations in caption of Table \ref{table:line_fits_typeC}.}
\label{table:line_fits_typeB}
\begin{tabular}{|l|l|l|l|l|l|l|}
\hline
Source        & Data points&  Fit slope  & Fit intercept  &Inclination                 & Sample & Ref.\\ \hline 
4U 1543-47       &2&  $0.123 \pm 0.0$ & $-0.259 \pm 0.0$  & $20.7 \pm 1.5 \degree$      & Low & [2]\\
GX 339-4         &2&  $0.031\pm 0.0$   & $0.435\pm 0.0$   & $40 \leq i \leq 60 \degree$ & Low & [4]\\ 
XTE J1817-330    &5&  $0.160\pm 0.620$ & $-0.473\pm 3.271$ & -                          & Low   & - \\
\hline

XTE J1859+226    &7&  $0.035\pm 0.072$& $0.367\pm 0.362$ & $\geq 60 \degree$           & - 		& [6]	\\ \hline
XTE J1550-564    &8&  $-0.024\pm 0.014$& $0.697\pm 0.082$ & $74.7 \pm 3.8 \degree$      & High 	& [7]\\ 
H1743-322        &16&  $0.038\pm 0.094$& $-0.165\pm 0.461$ & $75 \pm 3 \degree$          & High	& [9]	\\ 
\hline
\end{tabular}
\end{table*}

\bsp	
\label{lastpage}
\end{document}